\documentclass[journal]{IEEEtran}
\usepackage{cite}
\usepackage[pdftex]{graphicx}
\usepackage[cmex10]{amsmath}
\usepackage{array} %Alignment package
\usepackage{mdwmath} %Alignment package
\usepackage{mdwtab} %Alignment package
\usepackage{gensymb}
\usepackage{amssymb}
\usepackage[tight,footnotesize]{subfigure} %Allows the use of subfigures
\usepackage{comment}
\usepackage{cleveref}
\usepackage{mathtools}
\usepackage{stfloats}
\usepackage{esint}
\usepackage{url} % \url{my_url_here}.
\usepackage{xcolor}
\usepackage{color}
\newcommand{\change}[1]{#1}

\renewcommand\Im{\operatorname{Im}}
\renewcommand\Re{\operatorname{Re}}

\graphicspath{{pdf/}{jpg/}}
\DeclareGraphicsExtensions{.pdf,.jpg}

\begin{document}
\title{Planar Shielded-Loop Resonators}

\author{Brian~B.~Tierney,~\IEEEmembership{Student Member,~IEEE,}
        and~Anthony~Grbic,~\IEEEmembership{Member,~IEEE}% <-this % stops a space
\thanks{The authors are with the Radiation Laboratory at the Department
of Electrical Engineering and Computer Science, University of Michigan, Ann Arbor, MI, 48109-2122 USA (e-mail: btierney@umich.edu; agrbic@umich.edu).}}% <-this % stops a space

\markboth{To Be Published, February~2014}%
{Shell \MakeLowercase{\textit{et al.}}: Bare Demo of IEEEtran.cls for Journals}

%\markboth{IEEE Transactions on Antennas and Propagation, February~2014}%
%{Shell \MakeLowercase{\textit{et al.}}: Bare Demo of IEEEtran.cls for Journals}




\maketitle

\begin{abstract}
%\boldmath
The design and analysis of planar shielded-loop resonators for use in wireless non-radiative power transfer systems is presented. The difficulties associated with coaxial shielded-loop resonators for wireless power transfer are discussed and planar alternatives are proposed. The currents along these planar structures are analyzed and first-order design equations are presented in the form of a circuit model. In addition, the planar structures are simulated and fabricated. Planar shielded-loop resonators are compact and simple to fabricate. Moreover, they are well-suited for printed circuit board designs or integrated circuits.
\end{abstract}


% \fi
\IEEEpeerreviewmaketitle

\section{Introduction}

\begin{figure}[!t]
\centering
    \subfigure[A coaxial shielded-loop resonator. A loop is formed from a length of coaxial transmission line. The outer ground conductors are connected together. The outer conductor in this graphic is semitransparent, revealing that the inner conductor is open-circuited. A small piece of the outer conductor is removed halfway around the loop.]{
    \includegraphics[width=3.5in]{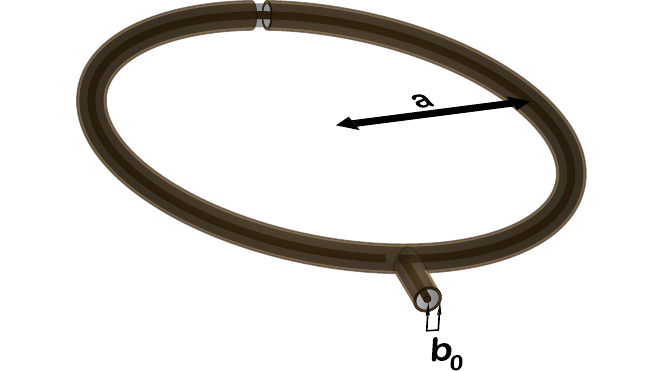}
    \label{fig:structureShieldedLoop}
    }
    \ \
    \subfigure[A planar, shielded-stripline loop resonator with a shifted slit in the ground conductor. The effect of shifting the slit is discussed in this paper.]{
    \includegraphics[width=3.5in]{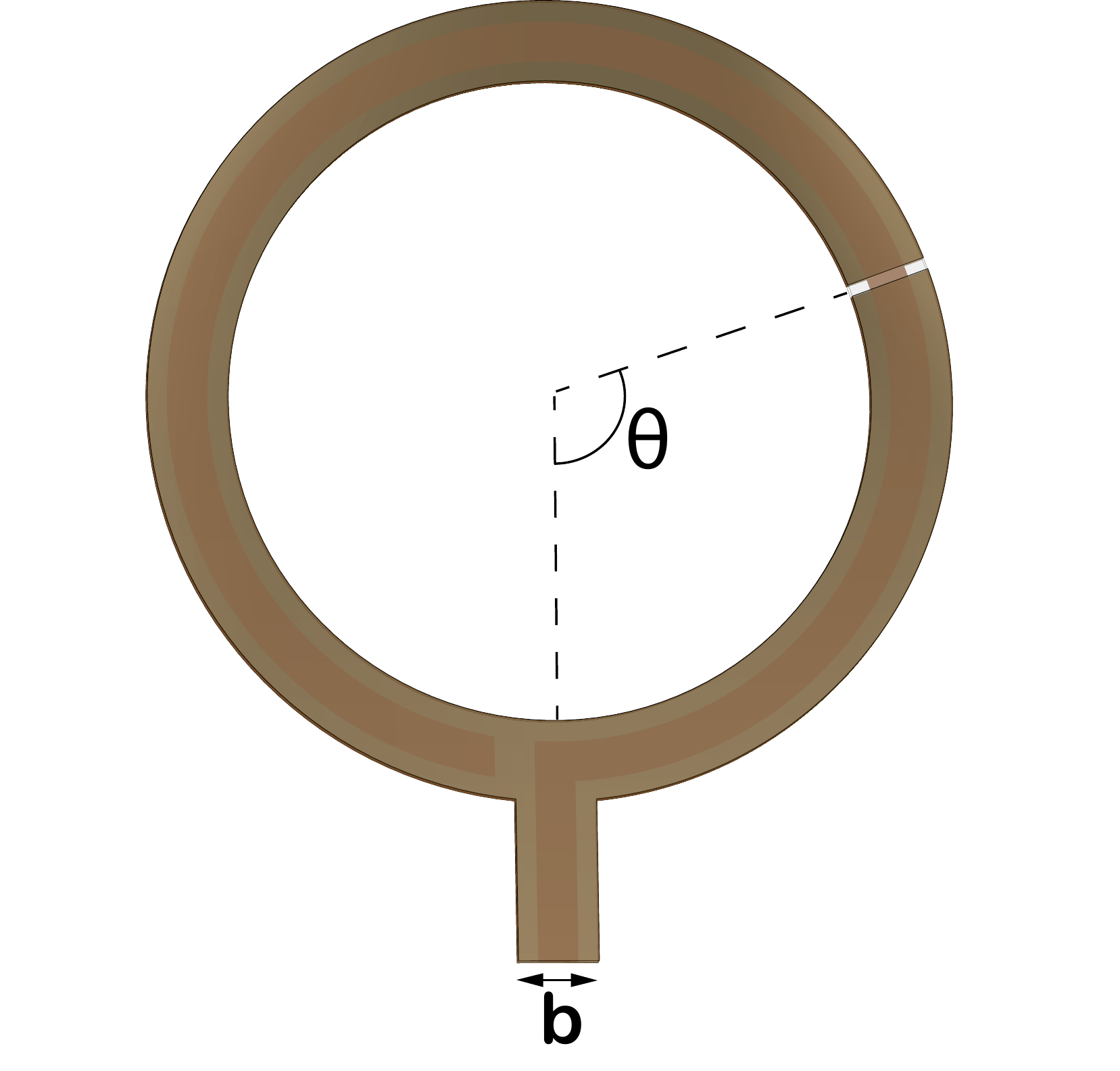}
    \label{fig:planarLoop}
    }
\caption{Coaxial and planar shielded-loop resonators.}\label{fig:shieldedLoops}
\end{figure}

\IEEEPARstart{T}{HE} wireless transfer of electromagnetic energy provides an often indispensable convenience when physical interconnects are bulky, dangerous, or simply not feasible. Inductive coupling methods are attractive because they are safe, affordable, and simple to implement. However, the coupling is quasistatic, which inherently limits operation to the near field. In 2007, wireless non-radiative power transfer (WNPT) was demonstrated at mid-range distances using resonant inductive coupling \cite{Kurs06072007}. Since that time, WNPT has received significant commercial interest and numerous applications have been pursued including the wireless powering of consumer electronics, electric vehicles \cite{5289747} and implantable medical devices \cite{4648077,5752354,4716667,4967822}. WNPT has also been advanced and demonstrated by various research groups \cite{5437250,5997544,4677400,5617728,5762318}.\par
Coaxial, shielded-loop resonators were recently used as receiving and transmitting loops in a WNPT system \cite{5562278,6148316}. \change{The fields generated by shielded-loop resonators are similar to those generated by classical resonators. However,} these compact, self-resonant structures offer low losses, an input that is simple to feed, and confined electric fields, which could otherwise couple to nearby dielectric objects. A similar resonator was subsequently pursued in \cite{MMCE:MMCE20603} for WNPT. An RLC circuit model was derived for these loops in \cite{6148316} using the properties of the coaxial transmission line from which they are constructed. Aside from power transfer, these resonant loops could be used for near-field communication \cite{968775,1639252}. Non-resonant versions have been used for magnetic-field probing \cite{1140518,1697139}.\par
The loops in \cite{6148316,MMCE:MMCE20603} were fabricated using lengths of semirigid coaxial transmission line. However, these types of loops can be difficult to fabricate. Moreover, their fabrication may not be repeatable. A planar structure made from a printed circuit board (PCB) would not only be simpler to fabricate but would also allow transmission-line properties, such as the characteristic impedance, $Z_0$, to be tailored. These resonators could be used to wirelessly power single chips \cite{4834116} or 3-D integrated circuits \cite{5751455,6148316,4115025}. The demand for chip real estate makes these compact, planar shielded-loop resonators advantageous. Planar shielded-loop resonators make economical use of space, since the inductance and capacitance are formed using the same conductive loops. This need for compact WNPT structures in practical systems is often stressed in literature \cite{6206305,6147134,6215789}.\par
In this paper, planar stripline and microstrip loop resonators are proposed and analyzed. In coaxial shielded loops, currents exist on the interior and exterior of the outer conductor, separated by the skin effect \cite{5562278,6148316}. However, this paper shows that such separation is not required. Equations for characteristic impedance $Z_0$, conductor attenuation $\alpha_c$, and dielectric attenuation $\alpha_d$ \cite{pozar,wheeler,hammerstadJensen} are used to develop first-order mathematical models for the stripline and microstrip resonators. The models are compared to both full-wave simulation and measurement. The trends of efficiency versus loop parameters are explored and compared for the different structures. The effect of re-positioning the slit in the ground conductor of the shielded planar loops is also explored.\par

\section{Background}\label{sec:theory}

Planar shielded-loop resonators behave in much the same way as their coaxial counterparts. Therefore, coaxial shielded loops are reviewed before introducing planar shielded-loop resonators. In addition, re-positioning the slit in the ground conductor is discussed.

\subsection{Coaxial Shielded-Loop Resonators}\label{subsec:coaxial}

\begin{comment}
\begin{figure}[!t]
\centering
\includegraphics[width=2.5in]{Coax_Full.pdf}
\caption{A coaxial shielded-loop resonator. A loop is formed from a length of coaxial transmission line. The outer ground conductors are connected together. The outer conductor in this graphic is semitransparent, revealing that the inner conductor is open-circuited. A small piece of the outer conductor is removed halfway around the loop.}
\label{fig:structureShieldedLoop}
\end{figure}
\end{comment}

\begin{figure}[!t]
\centering
\includegraphics[width=3.5in]{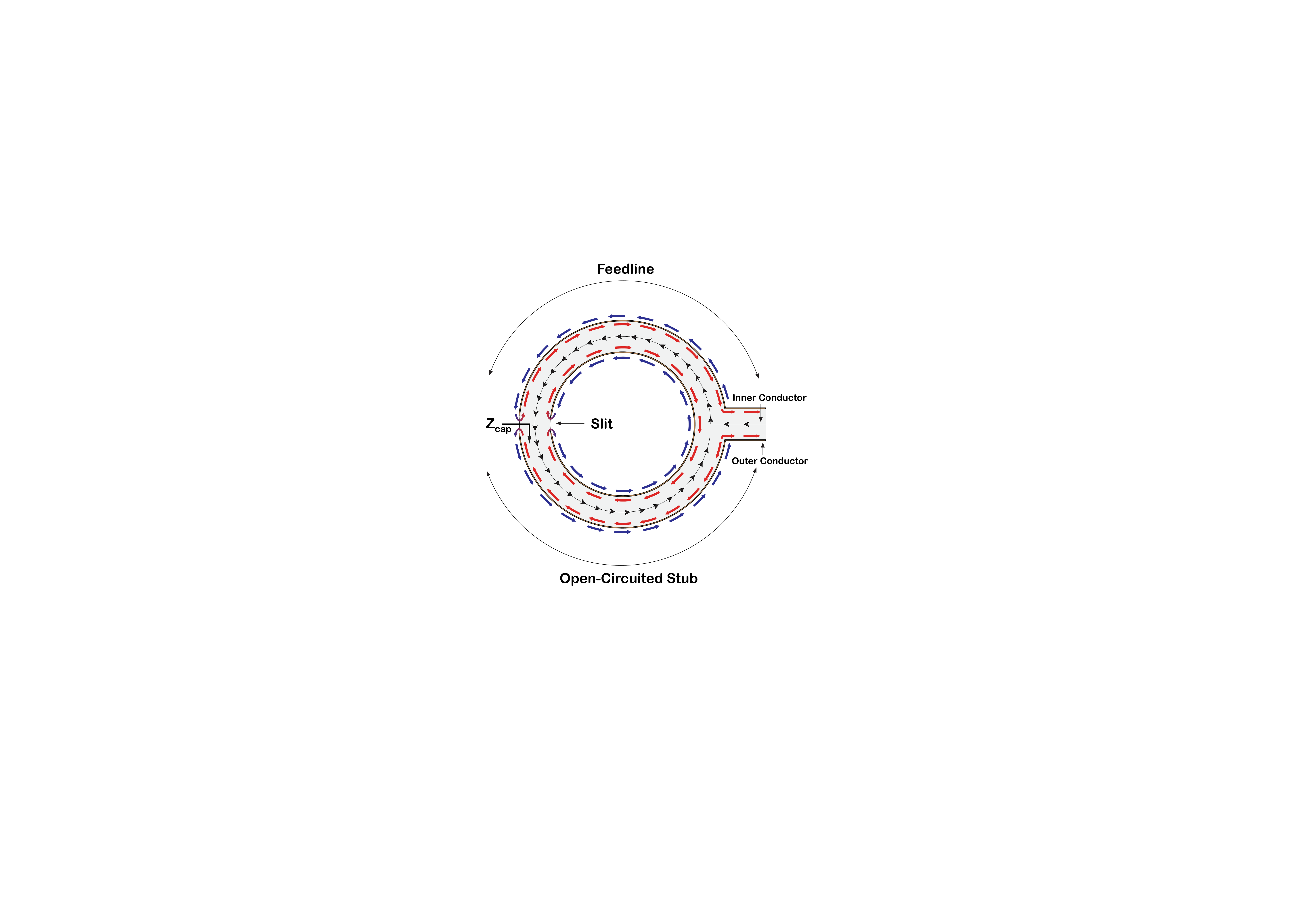}
\caption{The behavior of currents for a coaxial shielded-loop resonator. The current enters the input of the loop and propagates along the interior of the line to the slit in the outer shell. The ground current on the interior (red), wraps around to and propagates along the exterior of the structure (blue). This exterior current returns to the interior and ends in the open-circuited stub section.}
\label{fig:coaxialCurrentBehavior}
\end{figure}

\begin{figure*}[!t]
\centering
\includegraphics[width=7.0in]{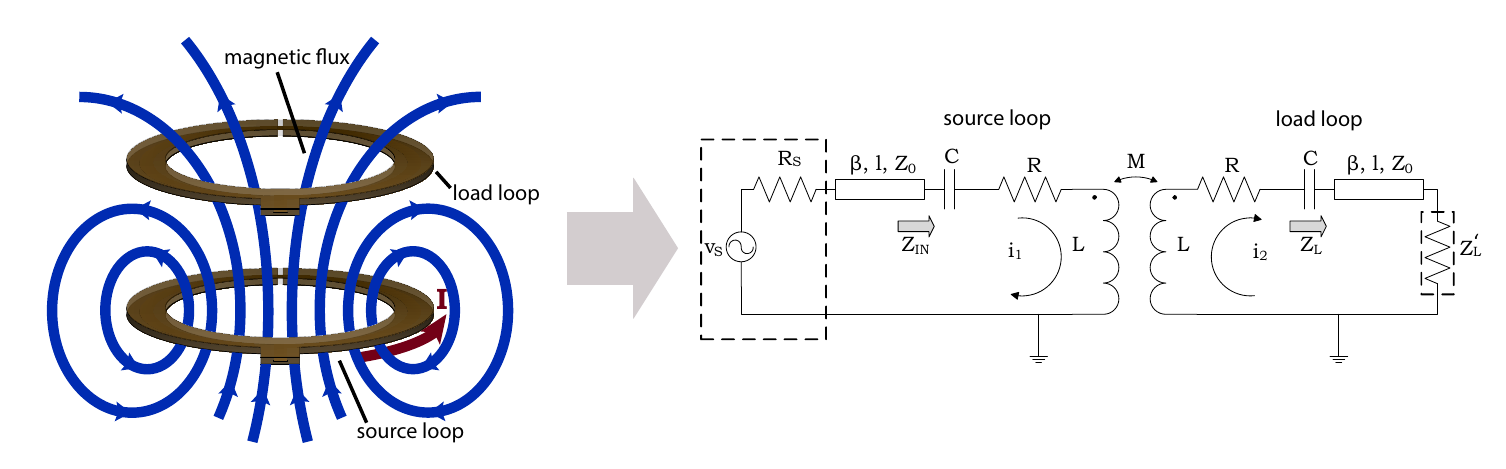}
\caption{Depiction of magnetic coupling between two identical loop resonators and the associated circuit model. The resonators shown are of the planar type presented in this paper. \change{The feedline transmission line is characterized by a propagation constant $\beta$, a physical length $l$, and a characteristic impedance $Z_0$. The input source is represented by input voltage $v_s$ and input impedance $R_s$.}}
\label{fig:basicCircuitModel}
\end{figure*}

Figure \ref{fig:structureShieldedLoop} shows a shielded-loop resonator constructed from a loop of coaxial transmission line with circular cross section. The outer conductor is continuous except for a small slit halfway around the loop. Figure \ref{fig:coaxialCurrentBehavior} shows the behavior of the current on a coaxial shielded-loop resonator. The current enters the input of the loop and propagates along the interior of the coaxial line to the slit in the outer ground conductor. It then wraps around the exterior of the loop to the opposite end of the slit, returning to the interior and ending at the open-circuited stub. The current does not traverse the slit directly due to the high reactance presented by the gap. As a result, the width of the slit does not appreciably affect the resonant frequency. If the outer conductor is thick (several skin depths), a current propagating on the interior of the outer conductor will be isolated from the current propagating on its exterior. This results in a loop current (inductor) in series with the open-circuited stub. The stub can be modeled as a capacitor for electrical lengths less than a quarter-wavelength ($\lambda/4$). Two magnetically-coupled, symmetric, coaxial shielded-loop resonators are modeled by the circuit shown in Fig. \ref{fig:basicCircuitModel}. The circuit includes the feedline, which precedes the slit. \change{The feedline transmission line is characterized by a propagation constant $\beta$, a physical length $l$, and a characteristic impedance $Z_0$. The parasitic resistance $R$ represents conductor, dielectric, and radiative losses. The input source is represented by input voltage $v_s$ and input impedance $R_s$.}\par

\begin{comment}
\begin{figure*}[!t]
\centering
\includegraphics[width=5in]{RLC_wFeedline.pdf}
\caption{Circuit model of magnetically-coupled, shielded-loop resonators. The resonators are RLC circuits with an input feedline. The loops are not necessarily identical.}
\label{fig:basicCircuitModel}
\end{figure*}

\begin{figure}[!t]
\centering
\includegraphics[width=3.0in]{CoupledLoops.pdf}
\caption{CoupledLoops}
\label{fig:coupledLoops}
\end{figure}
\end{comment}

\begin{comment}
\begin{figure*}[!t]
\centering
    \subfigure[Depiction of magnetic flux through a load loop. The resonators shown are of the planar type presented in this paper.]{
    \includegraphics[width=2.0in]{CoupledLoops.pdf}
    \label{fig:coupledLoops}
    }
    \subfigure[Circuit model of magnetically-coupled, shielded-loop resonators. The resonators are RLC circuits with an input feedline. The loops are not necessarily identical.]{
    \includegraphics[width=4in]{RLC_wFeedline.pdf}
    \label{fig:basicCircuitModel}
    }
\caption{Magnetically-coupled, shielded-loop resonators.}\label{fig:coupledSystem}
\end{figure*}
\end{comment}

\subsection{Circuit Model for Shielded Loops}\label{subsec:theModel}

As discussed in \cite{6148316}, shielded-loop resonators can be modeled as RLC resonators. For a coaxial cross section, the inductance of the loop can be approximated using the following well-known equation \cite{sengupta}:
\begin{equation}
\label{eq:inductance}L = \mu r \left [ \ln{( \frac{8 a}{b_0} )} - 1.75 \right ]
\end{equation}
Here, $a$ is the mean radius of the loop and $b_0$ is the radius of the conductor cross section. \change{For an electrically-small stub of length $l$ ($\beta l << 1$), the impedance $Z_{cap}$ looking into the open-circuited stub is}:
\begin{equation}
\label{eq:capacitance} Z_{cap} = \frac{-j}{\omega C} \approx jZ_0/(\beta l) = \frac{-j}{\omega C^{\prime} l}
\end{equation}
Therefore, the capacitance of the stub is simply:
\begin{equation}
\label{eq:capacitanceApprox} C = C^{\prime} l \quad , \quad \beta l << 1
\end{equation}
Here, $C^\prime$ is the per-unit-length capacitance of the transmission line. From $L$ and $C$, the resonant frequency can be determined:
\begin{equation}
\label{eq:resFreq} \omega_0 = \frac{1}{\sqrt{L C}}
\end{equation}\par
The resistance $R$ accounts for four loss mechanisms: radiative loss, exterior conductor loss, equivalent series resistance (ESR) of the capacitor, and feedline loss. The resistances representing these losses are given by the following expressions \cite{MMCE:MMCE20603}:
\begin{align}
R_{rad} &= 31170 \left (  \frac{\pi a^2}{\lambda^2} \right )^2 \label{eq:rRad} \\
R_{c} &= \frac{a}{A_S^\prime} \sqrt{\frac{f \mu \pi}{\pi \sigma}} \label{eq:rCond}\\
R_{ESR} &= Real[Z_0 \coth(\gamma \pi a )] \label{eq:rESR} \\
R_{feed} &= R^\prime \pi a \label{eq:rFeedline}
\end{align}
where $\lambda$ is the free-space wavelength at the frequency of operation, $\gamma$ is the complex propagation constant of the transmission line, $R^\prime$ is the per-unit-length resistance of the transmission line, and $A_S^\prime$ is the outer conductor perimeter for a cross-sectional slice of the loop. \change{For the coaxial shielded-loop resonator shown in Figure \ref{fig:structureShieldedLoop}, $A_S^\prime = 2 \pi b_0$.} The complex propagation constant is defined as in \cite{pozar}:

\begin{equation}
\label{eq:pozarGamma} \gamma = \sqrt{ (R^\prime + j \omega L^\prime)(G^\prime + j \omega C^\prime) }
\end{equation}
where $G^\prime$ and $L^\prime$ are the per-unit-length dielectric conductance and per-unit-length inductance, respectively. \change{For the circuit model in Figure \ref{fig:basicCircuitModel}, the feedline loss is included in $R$ and the propagation constant is purely real ($\beta$).}\par
In (6), a uniformly-distributed surface current is assumed, while (8) assumes that the feedline is half the circumference of the loop. For electrically-small loops, $R_{ESR}$ can be simplified to:
\begin{equation}
\label{eq:rESRSimple} R_{ESR} = Real \left [ \frac{1}{\pi a (G^\prime + j \omega C^\prime)} \right ]
\end{equation}
The $\pi a$ term in (7) and (10) results from the length of the open-circuited stub being half the loop circumference.\par
Knowing the transmission-line properties ($R^\prime$, $G^\prime$, $L^\prime$, $C^\prime$), the external geometry of the structure, and the resonant frequency $\omega_0$ using (4), the $Q$ of the resonator can be determined:
\begin{equation}
\label{eq:qFactor} Q(\omega) = \frac{\omega L}{R}
\end{equation}

These equations can also be applied to loop resonators made from planar transmission lines, which will be discussed next. However, these equations are only an approximation. A more accurate extraction of the resonator parameters can be performed through full-wave simulation. In this paper, equations (1) through (11) are referred to as the first-order design equations. For the planar loop resonators presented in Section \ref{sec:planar}, another equation is added to relate the width of a planar cross section to the radius of a coaxial cross section.

\subsection{Location of the Ground Slit}

\begin{comment}
\begin{figure}[!t]
\centering
\includegraphics[width=2.0in]{shiftedCut.pdf}
\caption{Example: A shielded stripline loop resonator with a shifted slit in the ground conductor.}
\label{fig:rotatedCut}
\end{figure}
\end{comment}

The shielded-loop resonators used previously for wireless non-radiative power transfer employed a slit halfway around the loop \cite{5562278}, as shown in Fig. \ref{fig:structureShieldedLoop}. For shielded-loops used in field probing, this location is chosen to minimize the probe's response to the electric field \cite{1140518,1697139}. However, this is of less concern for wireless power transfer applications. Furthermore, the arguments made thus far to justify the operation of these planar structures did not stipulate a specified location for the slit. Therefore, a structure with a slit 10\degree\ from input (see Fig. \ref{fig:planarLoop}) will be introduced. Reducing the length of the feedline is advantageous, since this improves the performance of the loop over a broad frequency range (see Appendix \ref{sec:appendix}).

\section{Planar Shielded-Loop Resonators}\label{sec:planar}

\begin{comment}
\begin{figure}[!t]
\centering
    \subfigure[The cross section for a planar analog to the coaxial, shielded-loop resonator using a shielded-stripline transmission line. The center conductor is a planar strip and the outer conductor is of rectangular shape.]{
    \includegraphics[width=2.5in]{StriplineCrossSection.pdf}
    \label{fig:shieldedStriplineCrossSection}
    }
    \ \
    \subfigure[The cross section for the unshielded shielded-loop resonator using a stripline transmission line.]{
    \includegraphics[width=2.5in]{StriplineCrossSection_unplated.pdf}
    \label{fig:shieldedStriplineCrossSection_unplated}
    }
\caption{Stripline cross sections for planar loop resonators.}\label{fig:lowZA}
\end{figure}
\end{comment}

\begin{figure}[!t]
\centering
    \subfigure[The cross section for a planar analog to the coaxial, shielded-loop resonator using a shielded-stripline transmission line. The center conductor is a planar strip and the outer conductor is of rectangular shape.]{
    \includegraphics[width=3.5in]{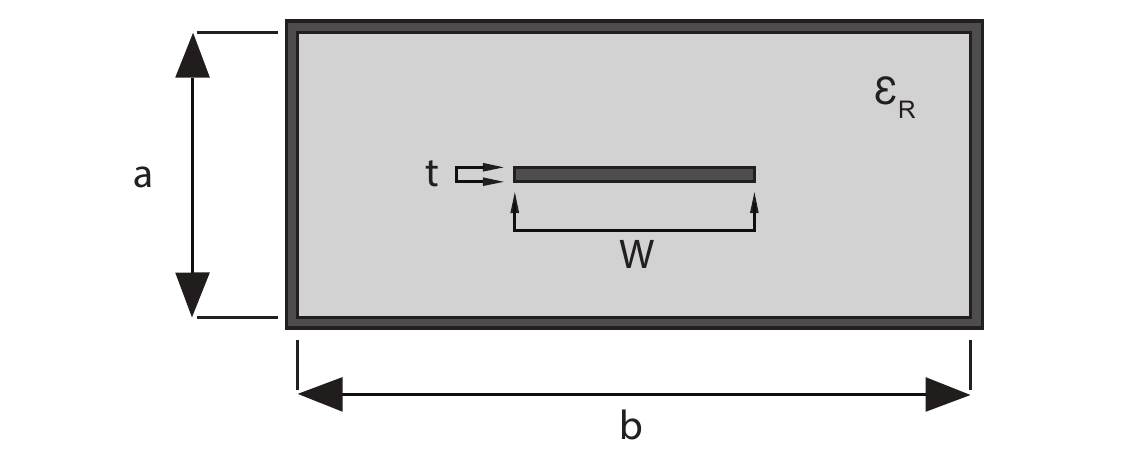}
    \label{fig:shieldedStriplineCrossSection}
    }
    \ \
    \subfigure[The cross section for the unshielded loop resonator using a stripline transmission line. The separation of currents are depicted in the structure. The loop current on the exterior now exists partially on the interior of the ground conductor. This behavior has been verified through full-wave simulation. The opposing currents on the ground conductor will separate.]{
    \includegraphics[width=3.5in]{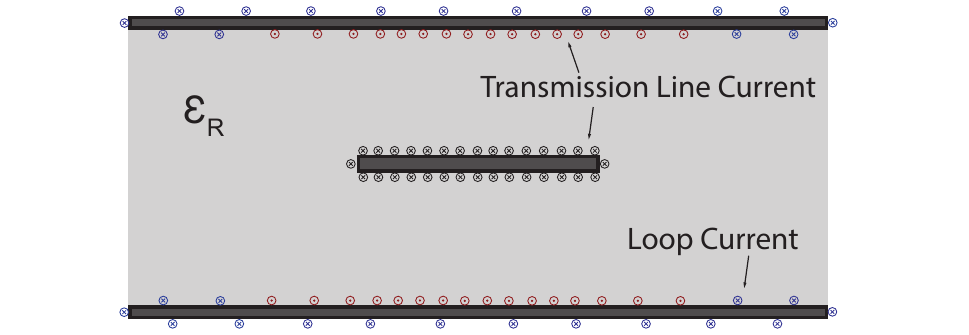}
    \label{fig:currentSeparation}
    }
\caption{Shielded and unshielded stripline cross sections for planar loop resonators.}\label{fig:striplineCrossSection}
\end{figure}

Coaxial, shielded-loop resonators (see Fig. \ref{fig:structureShieldedLoop}) can be difficult to fabricate, especially with repeatability. A planar structure, like that shown in Figure \ref{fig:planarLoop}, made from a printed circuit board is simpler to fabricate and allows tailoring of transmission-line properties, such as the characteristic impedance $Z_0$. The cross section of a planar shielded-loop resonator made from a shielded stripline is shown in Fig. \ref{fig:shieldedStriplineCrossSection}. To fabricate this structure, a loop can be cut from a PCB and its edges plated. Without plating, the dielectric on the edges of the loop is exposed, as shown in Figure \ref{fig:currentSeparation}, and the currents on the interior and exterior of the ground conductor are no longer physically isolated.\par
In Section \ref{sec:theory}, the behavior of the coaxial shielded-loop resonator was explained by the physical separation of the currents on the ground conductor. However, simulation shows that a stripline loop resonator behaves similarly whether it is shielded or not (Fig. \ref{fig:shieldedStriplineCrossSection} and \ref{fig:currentSeparation}, respectively). For unshielded structures, the loop current on the ground conductor naturally separates from the transmission-line current, as shown in Fig. \ref{fig:currentSeparation}. Therefore, these currents can be treated separately as a transmission-line current and a loop current. It will be shown that there are only minor differences in the extracted resonator parameters for loops having cross sections corresponding to Fig. \ref{fig:shieldedStriplineCrossSection} or Fig. \ref{fig:currentSeparation}.\par

\begin{comment}
\begin{figure}[!t]
\centering
\includegraphics[width=2.5in]{currentSeparation.pdf}
\caption{Depiction of current separation in the structure. The loop current on the exterior now exists partially on the interior of the ground conductor. This behavior is verified by full-wave simulation. The opposing currents on the ground conductor will separate.}
\label{fig:currentSeparation}
\end{figure}
\end{comment}

\begin{figure}[!t]
\centering
\includegraphics[width=2.5in]{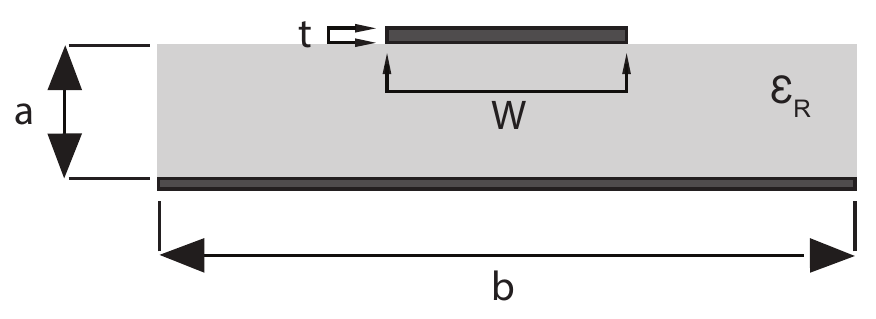}
\caption{The cross section for a loop resonator using a microstrip transmission line.}
\label{fig:MicrostripCrossSection}
\end{figure}

The cross section for an alternative loop resonator using a microstrip transmission line is shown in Fig. \ref{fig:MicrostripCrossSection}. With only two metal layers, this structure is simpler to fabricate but is not shielded.\par

 Analytically characterizing these planar structures is challenging. For example, the inductance is not solely due to the outer loop current, since the transmission line itself contributes to the overall inductance. However, a first-order approximation can be made with some simplifying assumptions. In particular, let's assume that the loop current flows uniformly on the exterior of the ground conductor. The formulas for resistive loss then become significantly simpler. The capacitance from the open-circuited stub can be easily approximated using (3). Assuming that the transmission line does not contribute to the total loop inductance, the inductance of a thin annulus can be used. This paper uses the following relationship:

 \begin{equation}
 \label{eq:inductanceDisk} b_0 = \frac{b}{4}
 \end{equation}
 where $b_0$ is the radius of the conductor cross section in equation (1) and $b$ is the width of planar cross section in the planar structure (see Fig. \ref{fig:shieldedStriplineCrossSection}). This relationship between round wires and strips can be found in \cite{tretyakov}. For each type of loop resonator, the properties of the transmission line and the geometry of the loop are all that are needed for this first-order analysis.

\section{Analytical Modeling}\label{sec:stripline}

In Section \ref{sec:theory}, the theory governing shielded-loop resonators was presented along with a corresponding RLC model. To determine the values of $R$, $L$, and $C$, the geometry and transmission-line properties of the resonators must be used. Therefore, design equations for both stripline and microstrip transmission lines are presented in this section. The resulting models will be subsequently compared to both simulation and experiment.

\subsection{Useful Transmission-Line Equations}

In this paper, Wheeler's formulas are used to compute the characteristic impedance $Z_0$ of a centered stripline \cite{wheeler}. The formulas from Hammerstad and Jensen are used to calculate the characteristic impedance $Z_0$ of a microstrip transmission line \cite{hammerstadJensen}.\par

The equations for $Z_0$ can also be used to calculate $R^\prime$, $G^\prime$, $C^\prime$, and $L^\prime$. Indeed, the conductor attenuation $\alpha_c$ is computed using Wheeler's incremental inductance rule, which is valid for any type of transmission line \cite{pozar}:

\begin{equation}
\label{eq:wheelerInductance} \alpha_c = \frac{R_s}{2 Z_0 \eta} \frac{d Z_0}{d l} \\
\end{equation}
where $l$ refers to the distance by which the walls of the conductors recede (see \cite{pozar}) and $R_s$  is the sheet resistance of the conducting surfaces, which can be derived from the frequency $f$ and conductivity $\sigma$ \cite{pozar}:

\begin{equation}
\label{eq:sheetResistance} R_s = \sqrt{ \frac{ f \mu \pi }{ \sigma } } \\
\end{equation}

Equation (14) assumes a uniform current distribution on the conductor's surface. However, due to the proximity effect this is only an approximation \cite{terman1943radio}. Therefore, the conductor attenuation $\alpha_c$ will be slightly higher than predicted by (13).\par
The dielectric attenuation $\alpha_d$ is computed using the following formula presented in \cite{pozar}, which is also valid for any type of transmission line:
\begin{equation}
\label{eq:dielectricAttenuation} \alpha_d = \frac{ k \tan{\delta} }{2} (Np/m)
\end{equation}\par
The per-unit-length resistance $R^\prime$ and conductance $G^\prime$ are also needed. From \cite{orfanidis}, these can be computed for a transmission line using $\alpha_c$ and $\alpha_d$, respectively:

\begin{subequations}
\label{eqs:RprimeandGprime}
\begin{alignat}{1}
R^\prime &= 2 \alpha_c Z_0 \\
G^\prime &= \frac{2 \alpha_d}{Z_0}
\end{alignat}
\end{subequations}

The per-unit-length capacitance $C^\prime$ and inductance $L^\prime$ are required as well. Since the transmission-line fields are quasi TEM:
\begin{subequations}
\label{eqs:LprimeandCprime}
\begin{alignat}{1}
v_p &= \frac{c}{\sqrt{\epsilon_{eff}}} \\
C^\prime &= \frac{1}{v_p Z_0} \\
L^\prime &= Z_0^2 C^\prime
\end{alignat}
\end{subequations}

Here, $v_p$ is the phase velocity of the wave and $\epsilon_{eff}$ is the effective permittivity of the transmission line. For stripline, $\epsilon_{eff} = \epsilon_{R}$, where $\epsilon_{R}$ is the relative permittivity of the dielectric.\par
\change{Using $R^\prime$, $G^\prime$, $L^\prime$, $C^\prime$, and $Z_0$ for the various geometries, the values of $R$, $L$, and $C$ for the circuit model (see Figure \ref{fig:basicCircuitModel}) are calculated}. The resulting $Q$ and resonant frequency $f_0$ for each geometry is compared to those extracted from full-wave simulation.

\section{Stripline Loop Resonators}

Stripline shielded-loop resonators are simulated using the commercial electromagnetic solver Ansys HFSS. To extract the resonator $R$, $L$, and $C$ parameters, the S-parameters are de-embedded to the location of the slit. Then, the de-embedded input impedances are used to extract the RLC parameters.\par

\subsection{Extraction Example}

To demonstrate the RLC extraction procedure for these structures, an example is provided here for a stripline shielded-loop resonators with an inner conductor width $W$ = 10 mm. From the full wave simulation, the characteristic impedance $Z_0$ and electrical length $\beta l$ of the feedline is found to be 17.6 $\Omega$ and 17.8\degree\, respectively, at 30 MHz. Using these numbers, the simulated input impedance $Z_{IN}(f)$ versus frequency $f$ is de-embedded to the location of the slit by first converting to S-parameters and applying a frequency-dependent phase shift:

\begin{align}
S_{11} &= \frac{Z_{IN}(f) - Z_0}{Z_{IN}(f) + Z_0} \\
S_{11}^\prime &= S_{11} e^{ j 2 \pi \frac{17.8}{180} \frac{f}{30} }
\end{align}

Here, $S_{11}^\prime$ denotes the de-embedded S-parameters, $f$ is in MHz, and $Z_0$ = 17.6 $\Omega$. The de-embedded input impedance $Z_{IN}^\prime(f)$ is given as:

\begin{align}
Z_{IN}^\prime(f) = Z_0 \frac{1 + S_{11}^\prime}{1 - S_{11}^\prime}
\end{align}

The resonant frequency $f_0$ is given by the frequency at which $\Im \{ Z_{IN}^\prime(f_0) \} = 0$. The input reactances at two distinct frequencies near $f_0$ is used to determine $L$ and $C$. The resonator resistance $R$ is simply the real part of the input impedance at resonance. The RLC parameters for the selected example are provided in Table \ref{table:shieldedStripline}.

\subsection{Shielded Stripline Resonator}
First, a set of shielded, stripline resonators with varying inner conductor widths are simulated to test the model presented in Section \ref{sec:theory}. Referring to the cross section in Fig. \ref{fig:shieldedStriplineCrossSection}, the properties of the simulated loop were:
\begin{itemize}
\item Conductor: copper ($\sigma$ = $5.8e7$ S/m)
\item Dielectric: Rogers RT/Duroid 5880
\item $\epsilon_R$: 2.2
\item $\tan \delta$: 0.009
\item Copper thickness ($t$): 70 $\mu$m
\item Loop radius ($r_0$): $9$ cm
\item Cross-sectional width ($b$):  $20$ mm
\item Cross-sectional thickness ($a$):  $3.32$ mm
\end{itemize}

\begin{table}[!t]
\centering
\caption{Parameters for shielded-stripline resonator from full-wave simulation.}
    \begin{tabular}{ | c | c | c | c | c | c |}
    \hline
      $W$  & $f_0 \ (MHz)$ & $L \ (\mu H)$ & $C \ (pF)$ & $R \ (\Omega)$ & $Q(f_0)$\\
    \hline
      2 mm & 54.7& 0.400& 21.2& 0.532&258\\
    \hline
      3 mm & 49.1& 0.379& 27.7& 0.394&297\\
    \hline
      4 mm & 44.9& 0.376& 33.5& 0.325&325\\
    \hline
      5 mm & 41.7& 0.367& 39.7& 0.279&345\\
    \hline
      6 mm & 39.2& 0.357& 46.4& 0.240&365\\
    \hline
      7 mm & 37.0& 0.350& 52.9& 0.215&378\\
    \hline
      8 mm & 35.2& 0.342& 59.9& 0.195&388\\
    \hline
      9 mm & 33.6& 0.344& 65.5& 0.183&396\\
    \hline
      10 mm & 32.2& 0.337& 72.6& 0.166&410\\
    \hline
    \end{tabular}
\label{table:shieldedStripline}
\end{table}
\ \par

From the extracted data, shown in Table \ref{table:shieldedStripline}, a few observations can be made. The resistance of the resonator decreases with increasing signal trace width $W$. Narrower widths exhibit higher loss due to increased current densities. Loop inductance also decreases with increasing signal width $W$. This is because the transmission line contributes to the overall inductance of the structure. This contribution decreases with increasing width $W$. Such a phenomenon can be seen even from (1), where the inductance of a loop of coaxial cross section decreases with increasing cross-sectional radius.\par
\begin{figure}[!t]
\centering
\includegraphics[width=3.5in]{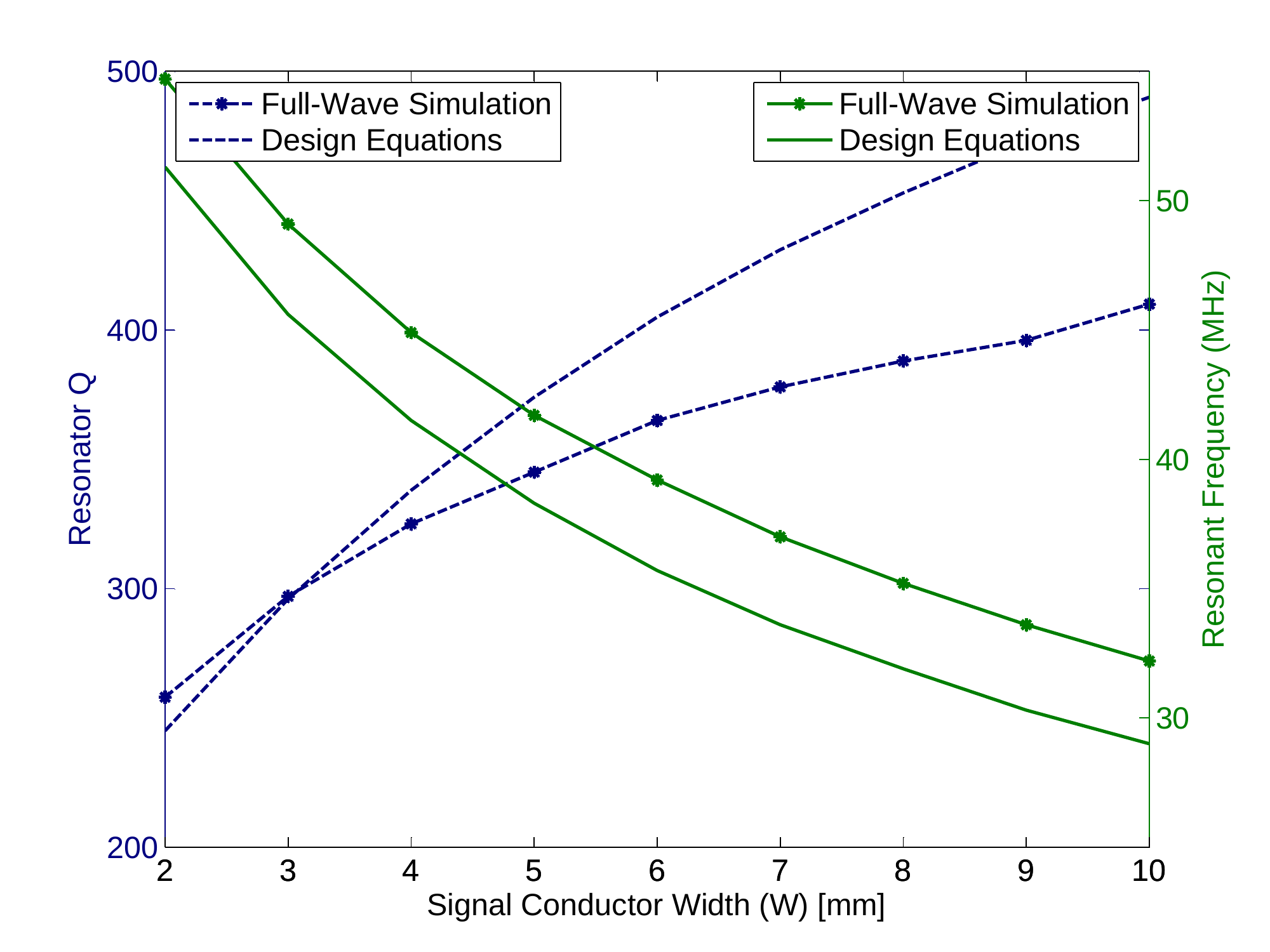}
\caption{The Q factor and resonant frequency of stripline shielded-loop resonators as a function of signal width W (see Figure \ref{fig:shieldedStriplineCrossSection}). Calculated results using first-order design equations are compared to those from simulation.}
\label{fig:striplinePlots}
\end{figure}

Figure \ref{fig:striplinePlots} shows the resonant frequency and Q factor extracted from full-wave simulation compared with those computed using the first-order design equations presented in Section \ref{sec:theory}. The computed resonant frequencies match fairly well with simulation and the Q factors exhibit the proper trend. The design equations generally overestimate the Q factor.\par

\begin{comment}
\begin{figure}[!t]
\centering
\includegraphics[width=3.5in]{striplineResFreq_B.pdf}
\caption{Shielded-stripline resonator resonant frequency: comparison between the first-order design equations and full-wave simulation.}
\label{fig:striplineResFreq}
\end{figure}

\begin{figure}[!t]
\centering
\includegraphics[width=3.5in]{striplineQ_B.pdf}
\caption{Shielded-stripline resonator Q factor: comparison between the first-order design equations and full-wave simulation.}
\label{fig:striplineQ}
\end{figure}
\end{comment}

\subsection{Unshielded Stripline Resonator}

 Next, a set of unshielded, stripline resonators (see Fig. \ref{fig:currentSeparation}) with varying inner conductor widths are simulated. The geometry and material properties are the same as for the plated case, but without metal plating on the edges.\par

\begin{table}[!t]
\centering
\caption{Parameters for unplated, shielded stripline resonator from full-wave simulation.}
    \begin{tabular}{ | c | c | c | c | c | c |}
    \hline
      $W$  & $f_0 \ (MHz)$ & $L \ (\mu H)$ & $C \ (pF)$ & $R \ (\Omega)$ & $Q(f_0)$\\
    \hline
      2 mm & 54.5& 0.406& 21.0& 0.527&264\\
    \hline
      3 mm & 49.1& 0.381& 27.6& 0.381&308\\
    \hline
      4 mm & 45.0& 0.372& 33.6& 0.343&307\\
    \hline
      5 mm & 41.8& 0.362& 40.0& 0.264&360\\
    \hline
      6 mm & 39.2& 0.357& 46.0& 0.231&382\\
    \hline
      7 mm & 37.1& 0.353& 52.2& 0.205&401\\
    \hline
      8 mm & 35.3& 0.344& 59.3& 0.183&415\\
    \hline
      9 mm & 33.7& 0.338& 66.1& 0.168&427\\
    \hline
      10 mm & 32.3& 0.336& 72.2& 0.155&440\\
    \hline
    \end{tabular}
\label{table:unplatedStripline}
\end{table}

From Table \ref{table:unplatedStripline}, the extracted resonator parameters for the unplated, stripline resonators are essentially the same as for the shielded-stripline resonators. However, the resistances of the unplated loops are slightly higher than for the plated loops. This occurs because the current is more spread out on the plated loop. Thus, plating the loops lowers the resistance but also adds a fabrication step.\par

\subsection{Shielded-Stripline Resonator with Shifted Slit}

As mentioned in Section \ref{sec:theory}, the slit in the ground conductor need not be placed opposite the input. Moving the slit toward the input feedline simply increases the length of the capacitive stub section. In addition, it decreases the length of the input feedline, which does not contribute to either the structure's overall inductance or capacitance, and can adversely affect the efficiency of a WNPT system (see Appendix \ref{sec:appendix}). Therefore, structures are also simulated for loops with the slit in the ground conductor 10\degree\ from the input feedline (see Fig. \ref{fig:planarLoop}). All other dimensions are the same as before. The extracted data is given in Table \ref{table:striplineShiftedCut}. From Table \ref{table:striplineShiftedCut}, the capacitance of the structure doubles, while the inductance remains roughly the same. Furthermore, the resistance of the structure decreases. Although the frequencies are lower, the Q of the structure at resonance is slightly larger due to the reduced loss.\par

\begin{table}[!t]
\centering
\caption{Parameters for shielded-stripline resonator with shifted ground slit from full-wave simulation.}
    \begin{tabular}{ | c | c | c | c | c | c |}
    \hline
      $W$  & $f_0 \ (MHz)$ & $L \ (\mu H)$ & $C \ (pF)$ & $R \ (\Omega)$ & $Q(f_0)$\\
    \hline
      2 mm & 37.8& 0.416& 42.5& 0.388&255\\
    \hline
      3 mm & 34.0& 0.396& 55.4& 0.302&280\\
    \hline
      4 mm & 31.2& 0.386& 67.5& 0.255&296\\
    \hline
      5 mm & 29.0& 0.368& 81.8& 0.219&306\\
    \hline
      6 mm & 27.3& 0.360& 94.5& 0.194&319\\
    \hline
      7 mm & 25.8& 0.351& 108.5& 0.175&326\\
    \hline
      8 mm & 24.5& 0.348& 121.1& 0.164&327\\
    \hline
      9 mm & 23.5& 0.338& 136.2& 0.147&339\\
    \hline
      10 mm & 22.5& 0.335& 149.0& 0.139&343\\
    \hline
    \end{tabular}
\label{table:striplineShiftedCut}
\end{table}

\subsection{Experimental Results}

\begin{figure}[!t]
\centering
\includegraphics[width=3.5in]{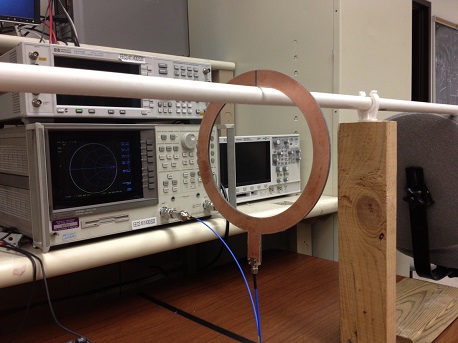}
\caption{A plated, shielded-stripline loop resonator. The radius of the loop is 9 cm.}
\label{fig:striplinePicture}
\end{figure}

Unplated stripline loops with a ground slit opposite the input were fabricated to compare with the simulation and the analytical model. Two Rogers RT/Duroid 5880 substrates were etched, bonded together using FR406 no-flow prepreg, and cut to form the loops. Figure \ref{fig:striplinePicture} shows the fabricated stripline loop resonator. The dimensions of the loop are the same as that for the loop described by Table \ref{table:unplatedStripline}. A trace width $W$ = 10 mm is used. The simulation and experimental results for this loop are shown in Table \ref{table:striplineMeasurements}.\par

\begin{table}[!t]
\centering
\caption{Comparison of simulation, experiment, and the first-order model for the unplated, stripline loop resonator ($W$ = 10 mm).}
    \begin{tabular}{ | c | c | c | c | c | c |}
    \hline
        & $f_0 \ (MHz)$ & $L \ (\mu H)$ & $C \ (pF)$ & $R \ (\Omega)$ & $Q(f_0)$\\
    \hline
      Simulation & 32.3& 0.336& 72.2& 0.16&440\\
    \hline
      Measurement & 32.1& 0.326& 75.4& 0.20&348\\
    \hline
      Model & 29.0& 0.364& 82.5& 0.14&490\\
    \hline
    \end{tabular}
\label{table:striplineMeasurements}
\end{table}

%\begin{figure}[!t]
%\centering
%\includegraphics[width=3.5in]{MultiturnResonator.jpg}
%\caption{A three-turn stripline resonator. The dimensions of this loop are the same as the %multi-turn simulated in the previous section.}
%\label{fig:multiturnPicture}
%\end{figure}

\section{Microstrip Loop Resonators}\label{sec:microstrip}
Microstrip loop resonators are simulated using the commercial electromagnetic solver HFSS. The resonator parameters $R$, $L$, and $C$ are determined using the methods presented in Section \ref{sec:stripline}.\par

\subsection{Simulation Results}

A set of planar, microstrip loop resonators with varying signal conductor widths are simulated to compare to the stripline loop resonators. The same dielectric and conductor as before are used. Referring to Fig. \ref{fig:MicrostripCrossSection}, the other properties of the loop were:
\begin{itemize}
\item $r_0$: $9$ cm
\item Cross-sectional width ($b$):  $20$ mm
\item Cross-sectional thickness ($a$):  $1.575$ mm
\end{itemize}

\begin{table}[!t]
\centering
\caption{Parameters for microstrip resonator from full-wave simulation.}
    \begin{tabular}{ | c | c | c | c | c | c |}
    \hline
      $W$  & $f_0 \ (MHz)$ & $L \ (\mu H)$ & $C \ (pF)$ & $R \ (\Omega)$ & $Q(f_0)$\\
    \hline
      2 mm & 68.6& 0.414& 13.0& 0.598&299\\
    \hline
      3 mm & 61.6& 0.409& 16.4& 0.477&331\\
    \hline
      4 mm & 57.1& 0.391& 19.9& 0.394&356\\
    \hline
      5 mm & 53.4& 0.381& 23.4& 0.343&372\\
    \hline
      6 mm & 50.9& 0.372& 26.4& 0.291&407\\
    \hline
      7 mm & 48.4& 0.362& 29.8& 0.256&430\\
    \hline
      8 mm & 46.0& 0.367& 32.6& 0.248&427\\
    \hline
      9 mm & 44.2& 0.364& 35.6& 0.229&442\\
    \hline
      10 mm & 42.8& 0.358& 38.7& 0.203&474\\
    \hline
    \end{tabular}
\label{table:microstrip}
\end{table}
\ \par

From Table \ref{table:microstrip}, it can be seen that the capacitance of the microstrip structure is lower than for the stripline structure. This follows from the decreased per-unit-length capacitance of the microstrip transmission line. The total inductance is increased because the height of the structure is decreased and the current on the ground conductor is more confined.\par
The resistance of the structure is lower compared to the stripline structure at similar frequencies. This is due to the fact that the fields partially exist in the air above the substrate, resulting in lower dielectric loss. However, this also means that the effective permittivity of the structure can be affected by objects in air directly above the substrate. Thus, the capacitance of the overall structure can be altered, which will detune the resonator. This is one disadvantage of the microstrip structure. On the other hand, fewer conductor layers are required and losses are lower.\par
\begin{figure}[!t]
\centering
\includegraphics[width=3.5in]{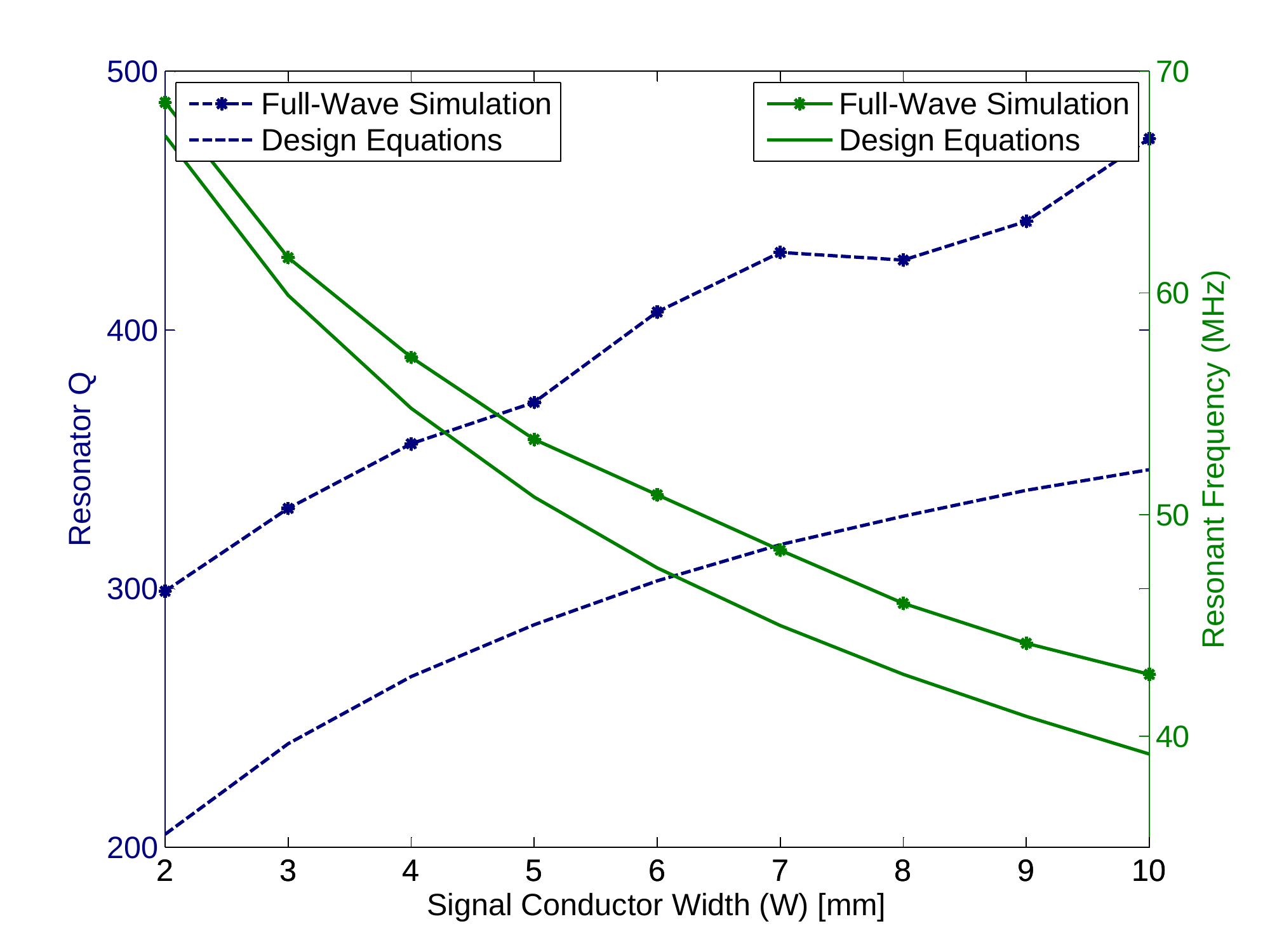}
\caption{The Q factor and resonant frequency of microstrip loop resonators as a function of signal width W (see Figure \ref{fig:MicrostripCrossSection}). Calculated results using first-order design equations are compared to those from simulation.}
\label{fig:microstripPlots}
\end{figure}

Figure \ref{fig:microstripPlots} shows the resonant frequency and Q factor extracted from full-wave simulation compared with those computed using the first-order model in Section \ref{sec:theory}. Again, the computed resonant frequencies match fairly well, and the Q factors exhibit the proper trend. The design equations, however, underestimate the Q factor.\par

\begin{comment}
\begin{figure}[!t]
\centering
\includegraphics[width=3.5in]{microstripResFreq_B.pdf}
\caption{Microstrip resonator resonant frequency: comparison between the first-order design equations and full-wave simulation.}
\label{fig:microstripResFreq}
\end{figure}

\begin{figure}[!t]
\centering
\includegraphics[width=3.5in]{microstripQ_B.pdf}
\caption{Microstrip resonator Q factor: comparison between the first-order design equations and full-wave simulation.}
\label{fig:microstripQ}
\end{figure}
\end{comment}

\subsection{Microstrip Resonator with Shifted Slit}

\begin{table}[!t]
\centering
\caption{Parameters for microstrip resonator with shifted ground slit from full-wave simulation.}
    \begin{tabular}{ | c | c | c | c | c | c |}
    \hline
      $W$  & $f_0 \ (MHz)$ & $L \ (\mu H)$ & $C \ (pF)$ & $R \ (\Omega)$ & $Q(f_0)$\\
    \hline
      2 mm & 46.4& 0.443& 26.5& 0.369&350\\
    \hline
      3 mm & 42.4& 0.431& 32.7& 0.289&398\\
    \hline
      4 mm & 39.3& 0.421& 39.0& 0.267&389\\
    \hline
      5 mm & 36.9& 0.407& 45.8& 0.234&403\\
    \hline
      6 mm & 34.8& 0.408& 51.3& 0.219&407\\
    \hline
      7 mm & 32.9& 0.401& 58.2& 0.209&398\\
    \hline
      8 mm & 31.5& 0.377& 67.6& 0.185&403\\
    \hline
      9 mm & 30.3& 0.384& 71.8& 0.179&409\\
    \hline
      10 mm & 29.3& 0.375& 79.0& 0.167&412\\
    \hline
    \end{tabular}
\label{table:microstripShifted}
\end{table}

As with the stripline structures, a microstrip loop resonator with the ground slit shifted 10\degree\ from the feed is also simulated. The extracted resonator parameters are shown in Table \ref{table:microstripShifted}. As before, the capacitances double while the inductances remain roughly the same. The Q factor of these resonators are similar to the original microstrip resonators described in the previous section, despite the lower frequencies.

\begin{comment}
\begin{table}[!t]
\centering
\caption{Parameters for three-turn, microstrip loop resonator.}
    \begin{tabular}{ | c | c | c | c | c | c |}
    \hline
      $d$  & $f_0 \ (MHz)$ & $L \ (\mu H)$ & $|S_{11}|_{\omega = \omega_0}$ & $\sum I \ (A)$ & $Q$\\
    \hline
      1 mm & 24.2 & 0.200& 0.943& 0.769&81\\
    \hline
      2 mm & 22.1 & 0.190& 0.966& 0.787&122\\
    \hline
      3 mm & 20.6 & 0.185& 0.974& 0.798&149\\
    \hline
      4 mm & 19.4 & 0.185& 0.979& 0.807&177\\
    \hline
    \end{tabular}
\label{table:threeturnMicrostrip}
\end{table}

\begin{table}[!t]
\centering
\caption{Parameters for two-turn, microstrip loop resonator.}
    \begin{tabular}{ | c | c | c | c | c | c |}
    \hline
      $d$  & $f_0 \ (MHz)$ & $L \ (\mu H)$ & $|S_{11}|_{\omega = \omega_0}$ & $\sum I \ (A)$ & $Q$\\
    \hline
      1 mm & 29 & 0.330& 0.952& 0.555&98.8\\
    \hline
      2 mm & 26 & 0.330& 0.968& 0.568&138.1\\
    \hline
      3 mm & 24.3 & 0.320& 0.976& 0.578&172.1\\
    \hline
      4 mm & 23 & 0.315& 0.979& 0.585&187.43\\
    \hline
    \end{tabular}
\label{table:twoturnMicrostrip}
\end{table}
\end{comment}

\subsection{Comparison with Lumped Matching}

\begin{table}[!t]
\centering
\caption{Parameters of RLC resonator using Lumped Matching.}
    \begin{tabular}{ | c | c | c | c | c |}
    \hline
      $f_0 \ (MHz)$ & $L \ (\mu H)$ & $C \ (pF)$ & $R \ (\Omega)$ & $Q(f_0)$\\
    \hline
      42.4 & 0.469 & 30 & 0.244 & 512\\
    \hline
    \end{tabular}
\label{table:lumped}
\end{table}

 Here, the performance of the non-shifted microstrip loop resonator of width $W$ = 10 mm is compared to the traditional lumped topology. To this end, an inductive loop is simulated using the commercial electromagnetic solver Ansys HFSS. The loop had radius $r_0$ = 9 cm and width 20 mm. The resulting inductance is $L$ = .469 $\mu$H and the parasitic resistance is $R$ = 0.184 $\Omega$. To match the resonant frequency $f_0$ = 42.9 MHz to that of the microstrip resonator, a 30 pF lumped capacitor is selected from Murata with equivalent series resistance of 0.06 $\Omega$. The result is an RLC resonator with characteristics given in Table \ref{table:lumped}.\par
 Although the Q and inductance of the lumped system is higher, achievable resonant frequencies are limited by by the availability of lumped capacitors. Moreover, the lumped capacitors limit the power-handling capabilities of the system.

\subsection{Experimental Results}

\begin{table}[!t]
\centering
\caption{Comparison of simulation, experiment, and the first-order model for the microstrip loop resonator ($W$ = 10 mm).}
    \begin{tabular}{ | c | c | c | c | c | c |}
    \hline
        & $f_0 \ (MHz)$ & $L \ (\mu H)$ & $C \ (pF)$ & $R \ (\Omega)$ & $Q(f_0)$\\
    \hline
      Simulation & 42.8& 0.358& 38.7& 0.20&474\\
    \hline
      Measurement & 42.1& 0.347& 41.3& 0.24&381\\
    \hline
      Model & 39.2& 0.364& 45.3& 0.26&346\\
    \hline
    \end{tabular}
\label{table:microstripMeasurements}
\end{table}

Microstrip loops with a ground slit opposite the input were fabricated to compare with simulations and the analytical model. The loops were milled and cut from a Rogers RT/Duroid 5880 substrate. The dimensions of the loop are the same as that for the loop described by Table \ref{table:microstrip}. A trace width $W$ = 10 mm is used. The simulation and experimental results for this loop are shown in Table \ref{table:microstripMeasurements}.\par

\subsection{Bandwidth}

\begin{figure}[!t]
\centering
\includegraphics[width=2.5in]{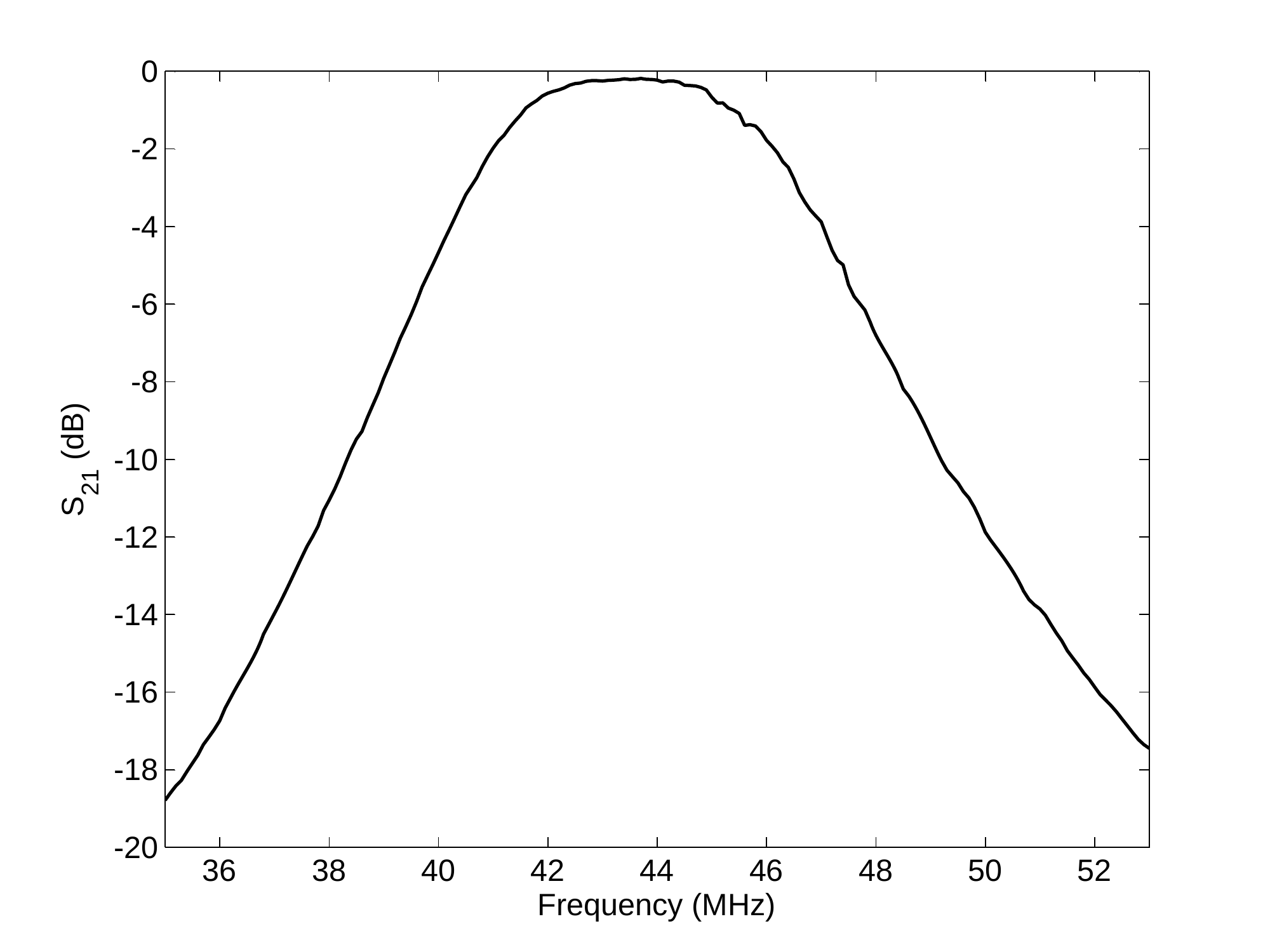}
\caption{Power transfer efficiency for two coupled microstrip loop resonators separated by 10 cm. A trace width $W$ = 10 mm and an ideal L-matching network are assumed.}
\label{fig:bandwidth}
\end{figure}

To explore the bandwidth of a system of coupled loops, the S-parameters of two microstrip loop resonators ($W = 10$ mm) separated by $10$ cm are measured. The power transfer efficiency $S_{21}$ is plotted in Figure \ref{fig:bandwidth} assuming an impedance match at $42.8$ MHz using an ideal L-matching network. The system achieves a power transfer efficiency of 96\% and a 3-dB bandwidth of 6 MHz. Since wireless power transfer systems typically employ continuous waves, this bandwidth is sufficient.

\section{Conclusion}

This paper introduced planar alternatives to coaxial shielded-loop resonators with a circular cross section, used previously for wireless non-radiative power transfer (WNPT). It was shown that physically shielding the transmission-line current from the inductive loop current is not required to maintain the resonant behavior of the structure. Therefore, the stripline loops' edges need not be plated. Structures with varying conductor widths were analyzed and simulated. Increased widths lowered both loss and the resonant frequency. For the same signal trace width $W$, the microstrip loops exhibited higher Q values than the stripline loops. Structures were fabricated and their measured values matched well with simulation. \par
These planar shielded loops offer a number of advantages over earlier resonant loops. First, the structures are inherently suitable for printed circuit boards and integrated circuits, since they are planar. Second, the structures are physically compact. Essentially, these structures are capacitors made from a pair of conductive annuli onto which a loop current (inductance) has been impressed. Third, the widths of the signal conductor can be easily varied to manipulate the loop parameters, such as the characteristic impedance. Fourth, the power-handling capabilites are superior to a simple loop with a lumped series capacitor. Finally, the fabrication variance for the capacitance of the structure is lower than that of a lumped series capacitor.\par
It was also shown in simulation that the location of the slit in the ground conductor of the proposed loops need not be directly opposite the input feedline. Instead, its location can be used to vary the capacitance and therefore the resonant frequency of the resonators. Moreover, a significant length of feedline can adversely affect the performance of the WNPT system (see Appendix \ref{sec:appendix}).\par
Future research directions include the use of air as a dielectric to create higher-Q loop resonators for wireless non-radiative power transfer. The dielectric loss tangent is a significant contributor to the parasitic loss of the structure, so an air dielectric would be advantageous. Other research directions include using an array of these planar structures to manipulate the near-field in an attempt to enhance mutual coupling between source and receiving loops.

\appendices

\section{Effect of Feedline on Power Transfer Efficiency}\label{sec:appendix}

\subsection{Effective Resistance of the Feedline}

\begin{figure}[!t]
\centering
    \subfigure[Schematic of an RLC resonator with an input feedline. Here the transmission line is characterized by its complex propagation constant $\gamma$, physical length $l$, and characteristic impedance $Z_0$.]{
    \includegraphics[width=3.5in]{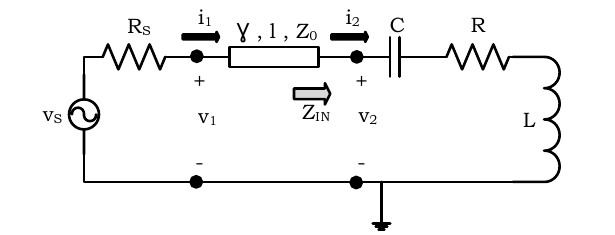}
    \label{fig:Proof_Schematic}
    }
    \ \
    \subfigure[Schematic of an RLC resonator with the power loss of the input feedline represented by an effective resistance $R_{EFF}$. A lossless transmission line is used to represent the phase delay.]{
    \includegraphics[width=3.5in]{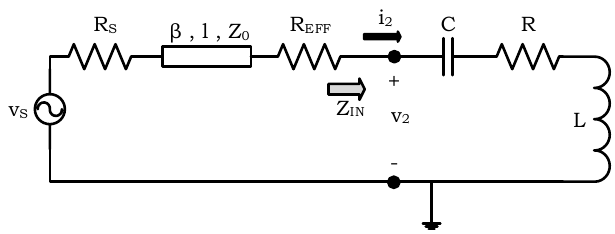}
    \label{fig:Proof_Schematic2}
    }
\caption{Effective feedline resistance $R_{EFF}$ for an RLC resonator with an input feedline.}\label{fig:lowZA}
\end{figure}

Consider an RLC resonator with a length of lossy transmission line attached to the input (see Figure \ref{fig:Proof_Schematic}), representative of a shielded-loop resonator. The transmission line can be decomposed into a lossless transmission line and an effective series resistance $R_{EFF}$ (see Figure \ref{fig:Proof_Schematic2}). The resistance $R_{EFF}$ can be determined by calculating the power $P_{feed}$ dissipated in the feedline and dividing it by $|I_2|^2$ (defined in Figure \ref{fig:Proof_Schematic2}):

\begin{equation}
R_{EFF} = \frac{2P_{feed}}{|I_2|^2}
\end{equation}

It can be shown that the effective resistance $R_{EFF}$ will reach a minimum at a frequency near the resonance of the RLC circuit. The effective resistance $R_{EFF}$ will increase as the excitation frequency is moved away from this minimum. This effect becomes more pronounced with longer transmission lines. For two magnetically-coupled RLC circuits (loops), the result is a decrease in maximum power transfer efficiency. This follows directly from the general equation for power transfer efficiency $\eta^\prime$ of the symmetric, coupled-loop system shown in Figure \ref{fig:basicCircuitModel} \cite{TierneyConference}:

\begin{align}
&\eta^\prime \triangleq \frac{P_{out}}{P_{source}} = \eta (1 - |\Gamma|^2) = \label{eq:efficiency}\\
&\frac{(1 - |\Gamma|^2)R_L (\omega M)^2}{R[ (R+R_L)^2 + (\omega L - \frac{1}{\omega C} + X_L)^2 ] + (\omega M)^2(R + R_L)} \nonumber
\end{align}
where $|\Gamma|^2$ is the power reflection coefficient, which depends on the impedance mismatch at the source, and $Z_L = R_L + j X_L$. Clearly, if $R$ increases, the efficiency $\eta$ of the system decreases. For loops that use a feedline, $R_{EFF}$ contributes to the total $R$.\par
To derive an expression for $R_{EFF}$ as a function of frequency, let's write an expression for the power lost in the feedline:
\begin{equation}
\label{eq:feedlinePowerLost} P_{feed} = \frac{1}{2} \left ( \Re \{ V_1 I_1^* \} - \Re \{V_2 I_2^* \} \right )
\end{equation}

From the equation for the impedance looking into a transmission line:

\begin{subequations}
\begin{alignat}{1}
\label{eq:v1i1} \frac{V_1}{I_1} &= Z_0 \frac{ Z_{IN} + Z_0 \tanh{ (\gamma l) } }{ Z_{0} + Z_{IN} \tanh{ (\gamma l) }} \\
\label{eq:v2i2} \frac{V_2}{I_2} &= Z_{IN}
\end{alignat}
\end{subequations}

Here, $\gamma = \alpha + j \beta$ is the complex propagation constant of the transmission line, where $\alpha$ quantifies the attenuation along the line. Using (24a) and (24b):

\begin{comment}
\begin{subequations}
\begin{eqnarray}
\label{eq:power1}&& \Re \{ V_1 I_1^* \} = \\
\nonumber &&|I_1|^2 Z_0 \frac{R_{IN} Z_0 ( 1 + p^2 + q^2)- q (R_{IN}^2 + X_{IN}^2+ Z_0^2)}{(Z_0 - q R_{IN}  - p X_{IN} )^2 + (p R_{IN}  - q X_{IN} )^2} \\
\nonumber && \triangleq |I_1|^2 A(R_{IN}, X_{IN})
\end{eqnarray}
\begin{equation}
\Re \{ V_2 I_2^* \} = |I_2|^2 R_{IN}
\end{equation}
\end{subequations}
\end{comment}

\begin{subequations}
\begin{alignat}{1}
&\Re \{ V_1 I_1^* \} = |I_1|^2 \Re \{ \frac{V_1}{I_1} \} = \label{eq:zin1} \\
&|I_1|^2 \frac{g(Z_{IN})}{( Z_0^{\prime} + p R_{IN} - q X_{IN} )^2 + ( Z_0^{\prime \prime} + q R_{IN} + p X_{IN} )^2} \nonumber \\ \ \nonumber \\
&\Re \{ V_2 I_2^* \} = |I_2|^2 R_{IN}
\end{alignat}
\end{subequations}

Here, $g(Z_{IN})$ is defined as:
\begin{equation}
\begin{aligned}
g(Z_{IN}) &=  R_{IN} |Z_0|^2 + |\tanh ( \gamma l)|^2 R_{IN} (Z_0^{\prime 2} - Z_0^{\prime \prime 2}) \label{eq:g} \\
&+ Z_0^{\prime} p ( |Z_{IN}|^2 + |Z_0|^2 ) + 2 Z_0^{\prime} Z_0^{\prime \prime} X_{IN} |\tanh ( \gamma l )|^2  \\
&- Z_0^{\prime \prime} q ( |Z_0|^2 - R_{IN}^2 - X_{IN}^2 )
\end{aligned}
\end{equation}

The impedance $Z_{IN}$, the characteristic impedance $Z_0$, and the function $\tanh{ (\gamma l) }$ have been decomposed into their real and imaginary parts so that $Z_{IN} = R_{IN} + j X_{IN}$, $Z_0 = Z_0^{\prime} + j Z_0^{\prime \prime}$, and $\tanh{ (\gamma l) } = p + j q$.

The power $P_{feed}$ dissipated in the feedline can be written as:

\begin{equation}
\label{eq:powerlost} P_{feed} = \frac{|I_2|^2}{2} \left ( \frac{|I_1|^2}{|I_2|^2} \Re \{ \frac{V_1}{I_1} \} - R_{IN}\right )
\end{equation}

Using (27), the effective resistance $R_{EFF}$ can be written as:

\begin{equation}
\label{eq:reff} R_{EFF} = \frac{2P_{feed}}{|I_2|^2} = \left [ \frac{|I_1|^2}{|I_2|^2} \Re \{ \frac{V_1}{I_1} \} - R_{IN} \right ]
\end{equation}

The current $I_1$ can be written in terms of $I_2$ using the Z-matrix $\bar{\bar{Z}}_{TL}$ of the transmission line:

\begin{equation}
\Longrightarrow \bar{\bar{Z}}_{TL} = Z_0\begin{bmatrix}
   \frac{ 1 + e^{- 2 \gamma l} }{ 1 - e^{- 2 \gamma l} } &  \frac{ 2 e^{- \gamma l} }{ 1 - e^{- 2 \gamma l} }  \label{eq:Zmatrix}  \\
   \frac{ 2 e^{- \gamma l} }{ 1 - e^{- 2 \gamma l} }  & \frac{ 1 + e^{- 2 \gamma l} }{ 1 - e^{- 2 \gamma l} }   \nonumber    \\
 \end{bmatrix}
\end{equation}

\begin{equation}
\nonumber \begin{bmatrix}
    V_1\\
    V_2\\
\end{bmatrix}
=
\begin{bmatrix}
    V_1 \\
    I_2 Z_{IN}\\
\end{bmatrix}
=
\bar{\bar{Z}}_{TL}
\begin{bmatrix}
    I_1 \\
    -I_2\\
\end{bmatrix}
\end{equation}

\begin{equation}
\label{eq:currentRelation} \Longrightarrow \frac{I_1}{I_2} = \left [  \frac{R_{IN} + j X_{IN}}{Z_0} \sinh (\gamma l) + \cosh (\gamma l)  \right ]
\end{equation}

Substituting (25a), (26), and (29) into (28) provides an equation for $R_{EFF}$ as a function of $\gamma$, $Z_0$, line length $l$, and $Z_{IN}$. Under the assumptions $Z_0^{\prime \prime} << Z_0^{\prime}$, $p << q$, and $R_{IN} \approx 0$, the equation simplifies to:

\begin{equation}
\boxed{
\label{eq:reff_simple} R_{EFF} = \frac{g(Z_{IN}) |\cosh (\gamma l)|^2}{|Z_0|^2}
}
\end{equation}

By taking the derivative $\delta / \delta X_{IN}$ of (30), it can be shown that for a given frequency and transmission line, $R_{EFF}$ achieves a minimum when $X_{IN} = X_{IN}^{min}$:

\begin{equation}
\label{eq:xmin}
\boxed{
X_{IN}^{min} = -\frac{Z_0^{\prime\prime} |\tanh ( \gamma l )|^2}{p + \frac{Z_0^{\prime \prime}}{|Z_0|} q}
}
\end{equation}

\begin{comment}
First, one should note that the sign of $q$ is negative. From \eqref{eq:complexTan}, $q$ can be written as:

 \begin{equation}
 \label{eq:imTan} \Im \{ \tan (\gamma l) \} = q = \frac{- \tanh{(\alpha l)} ( \tan^2 (\beta l) + 1)}{ 1 + \tan^2 (\beta l) \tanh^2 (\alpha l) }
 \end{equation}

From \eqref{eq:imTan}, if $\Im \{ \gamma l \} = -\alpha < 0$, then $q < 0$. Since the line is lossy, $-\alpha < 0$ and $q < 0$. From \eqref{eq:g} and \eqref{eq:zin1}, $\Re \{ \frac{V_1}{I_1} \}$ will increase as $|X_{IN}|$ increases (since $q < 0$). Moreover, from (\ref{eq:currentRelation}), the current ratio $\frac{|I_1|}{|I_2|}$ will increase as $|X_{IN}|$ increases, i.e. as the frequency moves from the resonant frequency. Therefore, by \eqref{eq:reff}, the effective resistance $R_{EFF}$. This effect will be more pronounced for longer feedlines.\par
\end{comment}

\begin{figure}[!t]
\centering
\includegraphics[width=3.5in]{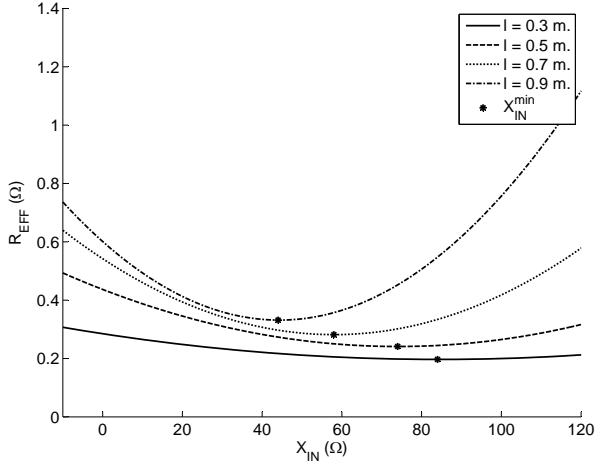}
\caption{The effective resistance $R_{EFF}$ for the transmission line.}
\label{fig:reff}
\end{figure}

Figure (\ref{fig:reff}) shows the variation of $R_{EFF}$ with $X_{IN}$ using (30) for a microstrip line of varying lengths attached to a lossless, reactive load $j X_{IN}$. The minima given by (31) are also plotted. The parameters of the microstrip, referring to Figure \ref{fig:MicrostripCrossSection}, are as follows:

\begin{itemize}
\item Height (a) = 813 $\mu$m %32 mil
\item Width (W) = 2 mm%78.58 mil
\item Relative Permittivity ($\epsilon_R$) = 3.0
\item Conductivity ($\sigma$) = 5.8e7 S/m
\item Conductor Thickness (t) = 35 $\mu$m
\item Loss Tangent ($\tan \delta$) = 0.001
\end{itemize}

The values of $\gamma$ and $Z_0$ are extracted from the commercial circuit solver ADS at 40 MHz:

\begin{align}
\gamma &= 0.0105 + j 1.31 \ m^{-1} \nonumber \\
Z_0 &= 50.38 - j0.3585 \ \Omega \nonumber
\end{align}

From Figure \ref{fig:reff}, a longer feedline exhibits a higher $R_{EFF}$ and a lower $X_{IN}^{min}$.\par
To verify (30), the analytical $|S_{11}|$ for the equivalent circuit of Figure \ref{fig:Proof_Schematic2} is compared with the simulated $|S_{11}|$ for the actual circuit of Figure \ref{fig:Proof_Schematic}.\par

\subsection{Implications for Coupled Loops}

\begin{comment}
\begin{figure*}[b]
\centering
\includegraphics[width=6.5in]{RLC_Schematic_wMatching.pdf}
\caption{Schematic of symmetric, magnetically-coupled WNPT system with matching networks.}
\label{fig:RLC_Schematic_wMatching}
\end{figure*}
\end{comment}

Referring to Figure \ref{fig:basicCircuitModel}, the input impedance $Z_{IN}$ for a symmetric system of magnetically-coupled loops in a WNPT system can be written as \cite{6148316}:

\begin{equation}
\label{eq:appendixInputImpedance} Z_{IN} = j( \omega L - \frac{1}{\omega C} ) + R + \frac{(\omega M)^2}{j( \omega L - \frac{1}{\omega C} ) + R + Z_L}
\end{equation}

For maximum power transfer \cite{6148316}:

\begin{equation}
\label{eq:optimalImpedance} Z_S = Z_L = j\left ( \frac{1}{\omega C} - \omega L \right ) + \sqrt{R^2 + (\omega M)^2}
\end{equation}

\begin{comment}
\begin{figure}[!t]
\centering
\includegraphics[width=3.5in]{Feedline_etaMax.pdf}
\caption{Maximum power transfer for a system of coupled shielded-stripline loop resonators like those characterized in Table \ref{table:shieldedStripline}. Here, the coupling distance is 10 cm. The feedline causes the maximum power transfer efficiency to drop off away from the maximum, for which the associated frequency depends on the parameters of the transmission line.}
\label{fig:Feedline_etaMax}
\end{figure}

\begin{figure}[!t]
\centering
\includegraphics[width=3.5in]{ShiftedMaxEta.pdf}
\caption{Maximum power transfer for a system of coupled shielded-stripline loop resonators like those characterized in Table \ref{table:striplineShiftedCut}. Here, the coupling distance is 10 cm. Because the feedline length is decreased, the maximum power transfer efficiency is greatly improved across the range of frequencies.}
\label{fig:ShiftedMaxEta}
\end{figure}
\end{comment}

\begin{figure}[!t]
\centering
    \subfigure[Maximum power transfer for a system of coupled shielded-stripline loop resonators like those characterized in Table \ref{table:shieldedStripline}. Here, the coupling distance is 10 cm. The feedline causes the maximum power transfer efficiency to drop off over frequency.]{
    \includegraphics[width=3.0in]{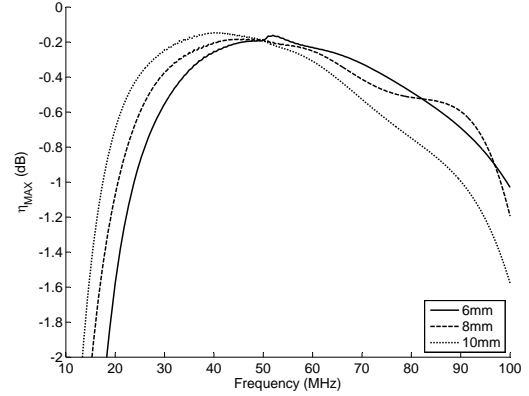}
    \label{fig:Feedline_etaMax}
    }
    \ \
    \subfigure[Maximum power transfer for a system of coupled shielded-stripline loop resonators like those shown in Table \ref{table:striplineShiftedCut}. The coupling distance is assumed to be 10 cm. The maximum power transfer efficiency is greatly improved across the range of frequencies by decreasing the feedline length.]{
    \includegraphics[width=3.0in]{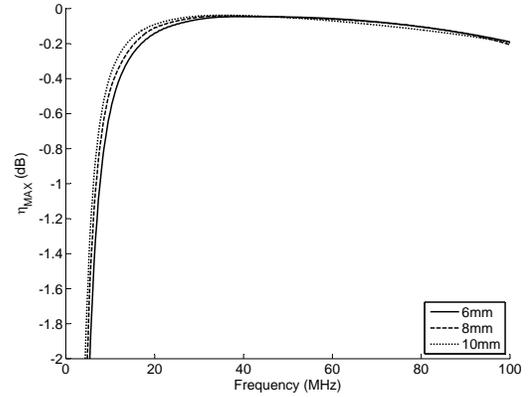}
    \label{fig:ShiftedMaxEta}
    }
\caption{Maximum power transfer efficiencies for different slit locations.}\label{fig:etaMax}
\end{figure}

%From \eqref{eq:appendixInputImpedance} and (\ref{eq:optimalImpedance}), it is clear that at the self-resonance of the loops, $( \frac{1}{\omega C} - \omega L ) = 0$, $Z_{IN}$ is purely real.
Using the same analysis as for a single loop, $R_{EFF}$ will be minimized at a point $X_{IN}^{min}$. As a result, the maximum achievable power transfer efficiency decreases as $|X_{IN} - X_{IN}^{min}|$ increases in the presence of an input feedline.\par
To verify these conclusions, pairs of coupled shielded-loop resonators from Tables \ref{table:shieldedStripline} and \ref{table:striplineShiftedCut} are simulated at a coupling distance of 10 cm. The maximum power transfer efficiency (achievable under a simultaneous conjugate match at both ports) is plotted versus frequency using the simulated S-parameters. Although thus far only the behavior of $R_{EFF}$ with respect to $X_{IN}$ has been discussed, the behavior versus frequency is similar. Figure \ref{fig:Feedline_etaMax} shows the maximum power transfer efficiency for three different systems of coupled shielded-stripline loop resonators like those presented in Table \ref{table:shieldedStripline}. Three different trace widths $W$ are used for the loops. It is clear that the efficiency is reduced for the higher frequencies. Figure \ref{fig:ShiftedMaxEta} shows the same results for the loops characterized in Table \ref{table:striplineShiftedCut} (shifted ground slit). The plots demonstrate that reducing the length of the feedline improves the performance of the loops away from the frequency corresponding to the minimum $R_{EFF}$.

\section*{Acknowledgment}
This work was supported by a Presidential Early Career Award for Scientists and Engineers (FA9550-09-1-0696).

\ifCLASSOPTIONcaptionsoff
  \newpage
\fi



%\bibliographystyle{IEEEtran}

%GATHER{referencePlanarShieldedLoops.bib}
\bibliographystyle{ieeetran}
\bibliography{referencePlanarShieldedLoops}

%\begin{biography}[{\includegraphics[width=1in,height=1.25in,clip,keepaspectratio]{mshell}}]{Michael Shell}

\begin{IEEEbiography}[{\includegraphics[width=1in,height=1.25in,clip,keepaspectratio]{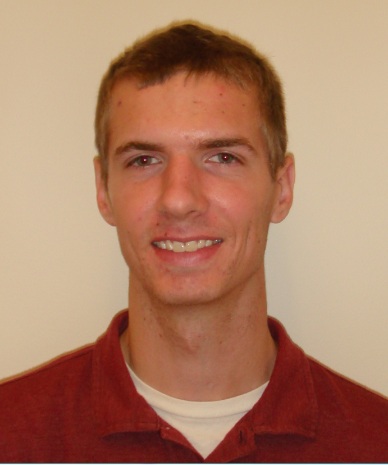}}]{Brian B. Tierney}
(S'11) received the B.S. degree in electrical engineering from Kansas State University, Manhattan, in 2011. He is currently working toward the Ph.D. degree in electrical engineering at the University of Michigan, Ann Arbor.\par
His research interests include wireless power transfer, metamaterials, leaky-wave antennas, tensor impedance surfaces, and analytical electromagnetics.
\end{IEEEbiography}

\begin{IEEEbiography}[{\includegraphics[width=1in,height=1.25in,clip,keepaspectratio]{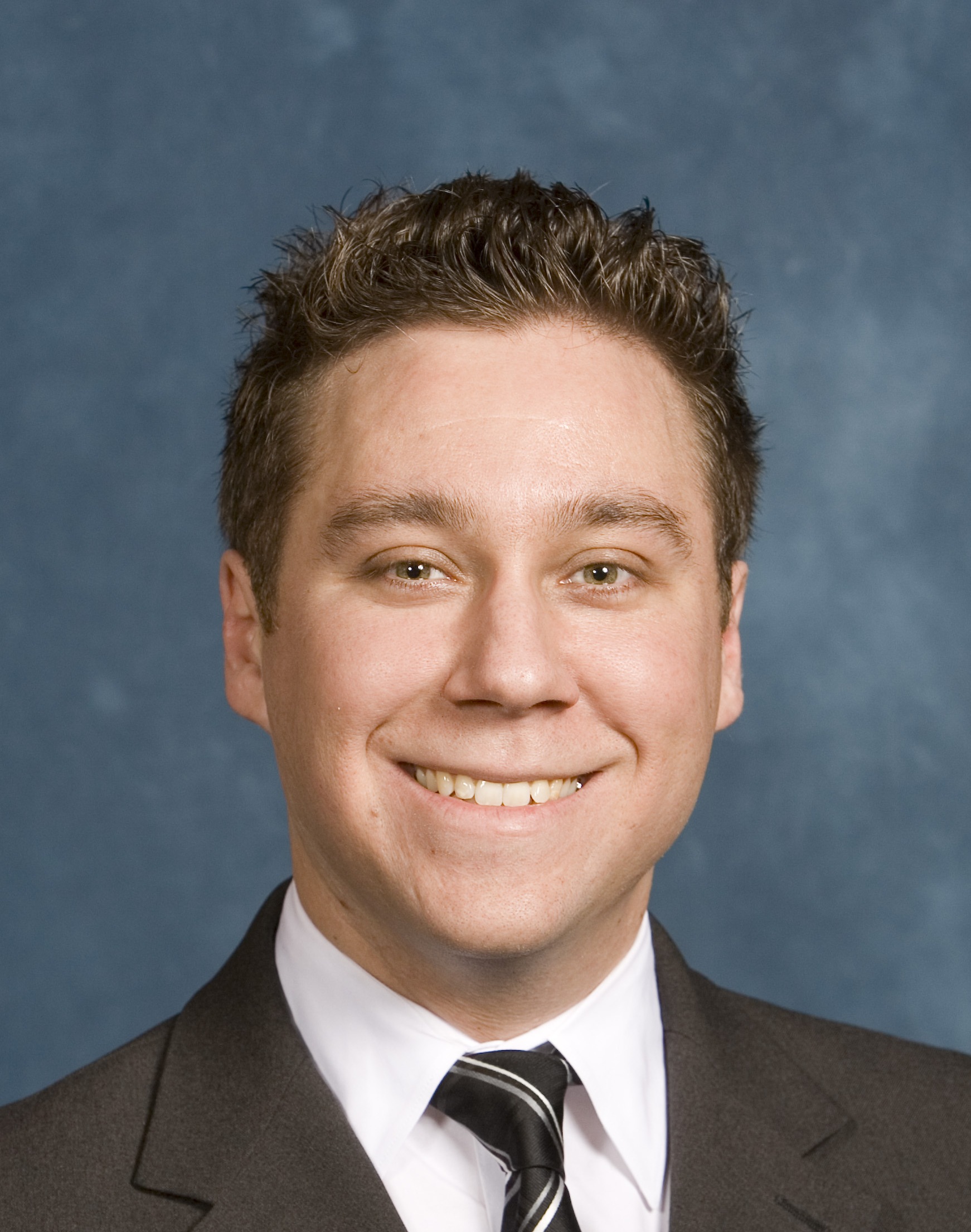}}]{Anthony Grbic}
(S'00–--M'06) received the B.A.Sc., M.A.Sc., and Ph.D. degrees in electrical engineering from the University of Toronto, Toronto, ON, Canada, in 1998, 2000, and 2005, respectively. In January 2006, he joined the Department of Electrical Engineering and Computer Science, University of Michigan, Ann Arbor, where he is currently an Associate Professor. His research interests include engineered electromagnetic structures (metamaterials, electromagnetic band-gap materials, frequency selective surfaces), antennas, microwave circuits, wireless power transmission systems, and analytical electromagnetics.\par
Dr. Grbic received an AFOSR Young Investigator Award as well as an NSF Faculty Early Career Development Award in 2008. In January 2010, he was awarded a Presidential Early Career Award for Scientists and Engineers. In 2011, he received an Outstanding Young Engineer Award from the IEEE Microwave Theory and Techniques Society, a Henry Russel Award from the University of Michigan, and a Booker Fellowship from the United States National Committee of the International Union of Radio Science (USNC/URSI). In 2012, he was the inaugural recipient of the Ernest and Bettine Kuh Distinguished Faculty Scholar Award in the Department of Electrical and Computer Science, University of Michigan. Anthony Grbic served as Technical Program Co-Chair for the 2012 IEEE International Symposium on Antennas and Propagation and USNC-URSI National Radio Science Meeting. He is currently the Vice Chair of AP-S Technical Activities, Trident Chapter, IEEE Southeastern Michigan section.
\end{IEEEbiography}

%\newpage


%\vfill


\end{document}